\documentstyle[12pt,preprint,tighten,prd,aps]{revtex}
\def\lsim{\mathrel{\lower2.5pt\vbox{\lineskip=0pt\baselineskip=0pt
          \hbox{$<$}\hbox{$\sim$}}}}
\def\gsim{\mathrel{\lower2.5pt\vbox{\lineskip=0pt\baselineskip=0pt
          \hbox{$>$}\hbox{$\sim$}}}} 
\begin{document}

\preprint{
\begin{tabular}{l}
gr-qc/9403008\\
Imperial/TP/93-94/20\\
Int. J. Mod. Phys. {\bf A10} (1995) 145
\end{tabular}
}
\title{Quantum gravity and minimum length}
\author{Luis J. Garay}
\address{Theoretical Physics Group, Blackett Laboratory,
Imperial College,\\
Prince Consort Road, London SW7 2BZ, U.K.}
\date{5 August 1994}
\maketitle
\begin{abstract}

The existence of a fundamental scale, a lower bound to any
output of a position measurement, seems to be a
model-independent feature of quantum gravity. In fact,
different approaches to this theory lead to  this result. The
key ingredients for the appearance of this minimum length are
quantum mechanics, special relativity and general relativity.
As a consequence, classical notions such as causality or
distance between events cannot be expected to be applicable at
this scale. They must be replaced by some other, yet unknown,
structure.

\end{abstract}

\section{Introduction}

Quantum gravity, the yet-to-be-built quantum theory of gravity,
involves a series of problems (see Ref. \cite{is93} for a
review) that have remained unsolved for many years. Most of the
problems arise from the fact that, unlike any other
interaction, gravity deals with the frame in which everything
takes place, with spacetime. Quantization of any interaction
except gravity leaves this frame unchanged. It is a passive
frame. But when gravity is brought onto the scene, the frame
itself becomes dynamical. It suffers the quantum fluctuations
of the other interactions and, even more, introduces its own
fluctuations. It becomes an active agent in the theory. Human
mind is used to putting everything into spacetime, so that it can
name and handle events. General relativity was a great change
in this sense. It made spacetime alive; but, although
dynamical, the relations between different events were still
sharply defined. Quantum mechanics changed this, too. In such a
sharply defined frame, objects became fuzzy; exact locations
were substituted by probabilities of finding an object in a
given region of space at a given instant of time.

The merging of quantum mechanics and special relativity gives
rise to the well known difficulties of dealing with
one-particle theories. Heisenberg's uncertainty principle and
the finiteness of the speed of light, when put together, lead
to creation and anihilation of particles if one tries to know
what is going on in a region within the  Compton wavelength of
the particle under study \cite{bl71}. The position-momentum
uncertainty principle $\Delta x\Delta p\gsim 1$ can be written
in terms of $x$, the position of a particle, and $E$, its
energy, as $\Delta x \Delta E\gsim 1$ for relativistic
particles (for which $E\sim p$). Therefore, if the position
uncertainty is smaller than a Compton wavelength, i.e. $\Delta
x  \lsim E^{-1}$, we get $E^{-1}\Delta E\gsim 1$ (throughout
this work we set $\hbar=c=1$, so  that the only dimensional
constant is Planck's length $l_*=\sqrt{\mbox{\small G}}$; also,
irrelevant numerical factors are ignored). Then, the
uncertainty in the energy of a particle is larger than its rest
mass and this makes the concept of particle unclear
\cite{he38}. We are facing  a resolution limit, a minimum
length for relativistic quantum  mechanics: it is not possible
to localize a particle with a better accuracy than its Compton
wavelength. In this case, it means that a multiparticle theory
is needed, i.e. quantum field theory. These problems are also
reflected when one tries to define a position operator. In
fact, an essentially unique position operator was defined by
Newton and Wigner \cite{nw49}. But this operator, whose
eigenvalues give the positions of a certain  particle, has the
following property: if a particle is localized in a certain
region at a certain time, then at any arbitrarily close instant
of time there is a non-zero probability of finding it anywhere;
therefore, the particle would be travelling faster than light.
This paradox disappears when, instead of dealing with a single
particle (only positive frequency modes are considered), one
allows the possibility of particle creation and anihilation
(i.e. negative frequency modes enter the theory).

The next step, from the theoretical point of view, and also in
the closely related energy scale, is to introduce gravity. Now
spacetime is dynamical. It is affected by, and also affects,
the objects and particles that it contains and that define it.
A quantum uncertainty in the position of a particle implies an
uncertainty in its momentum  and therefore, due to the
gravity-energy interaction, also implies an uncertainty in the
geometry, which in turn introduces an additional uncertainty in
position of the particle. Then, the expression for the position
uncertainty will contain (at least in some approximation
scheme) two pieces: the first one will correspond to the
standard Heisenberg principle and the second one to the
response of spacetime to the presence of this quantum
uncertainty. The geometry is now subject to quantum
fluctuations. What is the scale of these fluctuations?
\cite{wh64} To answer this question assume  that we want to
resolve a spherical region of radius $l$. We need a photon of
wavelength smaller than $l$; its energy will be greater than
$1/l$ and therefore we will be putting an energy density $\rho$
greater than $1/l^{4}$. Einstein's equations tell us that
$\partial^2 g \sim l_*^2 \rho\gsim l_*^2 /l^{4}$, where $l_*$
is Planck's length so that the gravitational potential (the
spacetime metric) generated by this photon is $g\gsim l_*^2
/l^{2}$ and therefore the length that is being measured will
have an uncertainty $\sqrt{g l^2}\gsim l_*$. Thus,
independently of any particular way of measuring the position,
the distance between two events will have a minimum
uncertainty  of the order of Planck's length. This conclusion,
which has been reached by means of rather crude arguments, can
be obtained from a variety of more sophisticated
thought-experiments and theoretical analyses of quantum gravity.

The existence of this fundamental length at Planck's scale
beyond which the very concepts of space and time lose their
meaning may have a similar meaning to the speed limit defined
by the speed of light in special relativity, as pointed out by
Markov \cite{ma80,ma81} (as quoted in Ref. \cite{bt88}): there
is no  way of going beyond this  border and its existence may
be inferred through the relativistic corrections to the
Newtonian theory. This would mean that a quantum theory of
gravity could be constructed only in ``this side of Planck's
border''. In fact, as we will see below, this analogy between
quantum gravity and special relativity is quite close: in the
latter you can accelerate forever even though you will never
reach the speed of light; in the former, given a coordinate
frame, you can reduce the coordinate distance between two
events as much as you want even though the proper distance
between them will not decrease beyond Planck's length.

The purpose of this paper is to review several scenarios in
which a minimum length arises in the context of quantum gravity
and also discuss some of the consequences. In what follows, we
will be concerned with uncertainties due to the quantum
fluctuations of the gravitational field itself or, in other
words, the fluctuations that are introduced when a measurement
is performed. However, there is another source of uncertainty
that will not be considered here: the quantum fluctuations of
the gravitational field due to the Heisenberg uncertainty
relations that affect the source of the gravitational  field
\cite{ps86,dd86,kf93}. This uncertainty decreases with
distance  to the source and it vanishes for sufficiently large
distances.

\section{Minimum length}

A minimum length at Planck's scale can be obtained from the
study of some thought-experiments and approaches to quantum
gravity. We describe some of them in this section.

\subsection{Microscope thought-experiments}

 The process of detecting  a particle with a microscope and
some related thought-experiments were carefully analyzed  by
Mead \cite{me64}. He gave a variety of treatments from the
simplest in which gravity was regarded as Newtonian, to the most
sophisticated in which general relativity was fully
implemented. For simplicity, we will  reproduce the Newtonian
argument.

Apart from the uncertainty due to the lack of knowledge of the
exact direction of the photon scattered by the observed
particle, $\Delta x\gsim 1/\Delta p$, there is another piece
that contributes to this uncertainty. This is due to the
gravitational interaction between the observed particle and the
photon. The particle will be attracted towards the photon with
an acceleration given by the Newtonian gravitational field
$l_*^2\omega/r^2$, where $\omega$ is the energy of the photon
and $r$ is the radius of the region of strong interaction (i.e.
during the time $r$ that the photon is within this region,
it cannot be focussed). During this time, the particle acquires
a velocity $l_*^2\omega/r$ and therefore travels a distance
$l_*^2\omega$ in the direction in which the photon is moving,
which is unknown. Projecting down to the $x$-axis, we find
$\Delta x\gsim l_*^2\Delta p$. Combining both pieces of
uncertainty, we get
\[
\Delta x\gsim {\rm max}\left(\frac{1}{\Delta p}, l_*^2\Delta
p\right)\gsim l_*.
\]
The physical explanation seems evident from this expression. To
get a high resolution you need high energy photons (due to the
standard Heisenberg's uncertainty relation), but the higher the
energy of the photons, the higher the gravitational interaction
and therefore the higher the disturbance. If the energy is
too high, gravitational effects will seriously affect the
particle and spoil the advantages  of using high energy photons.

Mead also showed that there is a limit in the time resolution
when synchronizing clocks (see also Ref. \cite{nit93}). We will
make use of the full general relativity theory (rather than the
Newtonian limit). In fact, the argument given above, although
qualitatively correct,  is not valid since the Newtonian
approximation breaks down at the scales that we are working.
Assume that we want to synchronize a clock with another
standard clock. We will do this by interchanging photons. The
reading of our clock has at least two sources of uncertainty.
The first one is due to the Heisenberg time-energy uncertainty
relation, which constrains the accuracy in the emission or
absorption of the photon, and therefore the reading of our
clock, to be smaller than $1/\Delta\omega$. The second source
of uncertainty is the gravitational interaction between the
photon and the clock. Assume that the clock and the photon
strongly interact in a region of radius $r$, and therefore
during a period $r$ (provided that the clock remains
stationary). The duration of this interaction, according to the
clock, will be $\sqrt{g_{00}}r$, where $g_{00}$ is the
time-time component of the gravitational field generated by the
photon (and suffered by the clock) and that can be seen to have
the form
\[
g_{00}=1-\frac{4l_*^2\omega}{r}.
\]
The uncertainty in the duration of the interaction due to the
frequency spread $\Delta \omega$ is
\[
\frac{2l_*^2\Delta\omega}{\sqrt{1-4l_*^2\omega/r}}\gsim
2l_*^2\Delta \omega.
\]
The uncertainty in the reading of our clock will then be
greater than both uncertainties:
\[
\Delta t\gsim {\rm max}\left( \frac{1}{\Delta \omega},
l_*^2\Delta \omega\right) \gsim l_*.
\]

It should be mentioned that these uncertainty relations are
low-energy approximations to what the full theory of quantum
gravity would give. Indeed, they are made up of two pieces: the
first one is due to the Heisenberg uncertainty principle and
the second is due to the dynamical response of spacetime to
such uncertainty. But this alteration should induce a new
uncertainty in the gravitational field, which would again add
an additional uncertainty, and so on. This whole series of
terms is an expression of the nonlinearity of the gravitational
interaction or, in other words, of the equivalence principle.

\subsection{Path-integral quantum gravity}

{}From different points of view, both Padmanabhan
\cite{pa85a,pa85b,pa86,pa87} and Greensite \cite{gr91} inferred
the existence of  a minimum length within a path integral
formulation of  quantum gravity; namely, the former worked in a
theory in which the conformal factor is quantized \cite{np83}
and the latter in lattice quantum gravity (see also Refs.
\cite{nbk93a,nbk93b,nff84}). They both reached the result that
the coincidence limit, in which the difference between
coordinates of two events in a given frame or the separation
between lattice points goes to zero, of the proper interval
between them is not zero but Planck's length. However, there
are some discrepancies as to the origin of this residual length
as we will see. Let us review their arguments, beginning with
Padmanabhan's approach.

For the sake of simplicity, we will restrict our attention to
the case of conformally flat metrics
\[
g_{\mu\nu} (x)=\left[ 1+\phi(x)^2\right] \eta_{\mu\nu},
\]
although the same result is obtained if we remove this
restriction. The path integral over the conformal fluctuation
$\phi$
\[
\int {\cal D}\phi\exp\left(\frac{-i}{l_*^2}\int d^4x
\eta^{\mu\nu}\nabla_\mu\phi\nabla_\nu\phi\right)
\]
can be evaluated in a closed form. This allowed Padmanabhan to
write the probability amplitude for a measurement of the
conformal fluctuation  to give a value $\phi$:
\[
{\cal A}(\phi)=\left(\frac{l}{l_*}\right)^{1/4}
\exp\left(-\frac{l^2}{l_*^2}\phi^2\right),
\]
where $l$ is the resolution of the measuring apparatus (the
radius of the region over which it averages). It should be
noted that if the region over which one measures is very large
($l\gg l_*)$, then the probability amplitude $\cal A$ will have
a sharp peak at flat spacetime (i.e. $\phi=0$). This
distribution provides a kind of uncertainty relation
\[
\Delta\phi\ l\gsim l_*.
\]
The greater the resolution (i.e. the smaller $l$), the greater
the fluctuations of the conformal factor will be.

Let us now calculate the coincidence limit for two close events
of the vacuum expectation value of its proper interval:
\[
\lim_{x\rightarrow y}\langle l^2(x,y)\rangle=\lim_{x\rightarrow y}
\left[ 1+\langle\phi(x)\phi(y)\rangle\right] l_{0}^2(x,y)
\]
where $l_{0}^2(x,y)$ is the proper interval in the metric
$\eta_{\mu\nu}$. However, from the form of the action for the
field $\phi$ (it is that of a free scalar field with the
``wrong sign'') it is straightforward to see that
\[
\langle\phi(x)\phi(y)\rangle=\frac{l_*^2}{l_{0}^2(x,y)},
\]
so that $l_*$, Planck's length,  is the minimum proper interval
for any two events.

Greensite \cite{gr91}, on the other hand, considered  linearized
gravity in a cubic lattice of points that represents spacetime.
Again, the arguments can be extended to full general relativity
without any change in the final result. Also, the tetrad
formulation is chosen in order to prevent configurations
with det$(-g)<0$ from giving any contribution to the path integral.
Then the gravitational field is represented by the tetrad
\[
e_\mu^a=\delta_\mu^a+l_*^2b_\mu^a
\]
and the path integration is performed over the field $b_\mu^a$.
The proper distance separation between two neighbouring events
in the lattice (in the $x$-direction, for instance) is
\[
\langle ds^2\rangle=\left[ 1+l_*^2\delta_{ab}\langle
b_1^a(x)b_1^b(x)\rangle\right] \epsilon^2,
\]
where $\epsilon$ is the lattice parameter. As above,
\[
\delta_{ab}\langle b_1^a(x)b_1^b(x)\rangle\sim \epsilon^2
\]
and therefore,
\[
\lim_{\epsilon\rightarrow 0}\sqrt{\langle ds^2\rangle}\sim l_*.
\]
As Greensite pointed out, in both approaches the fact that
configurations with negative det$(-g)$ are precluded shifts the
expectation value of $g_{\mu\nu}$, although the origin of this
shift is different: fluctuations of the conformal factor in
Padmanabhan's approach and all components of the metric in
Greensite's (in the latter the conformal fluctuations tend to
depress this effect rather than enhance it). There is also
another difference between both approaches. In the lattice, the
physical separation between two points separated by a finite
coordinate distance is infinite in the $\epsilon\rightarrow 0$
limit, while in Padmanabhan's approach, it is finite and
regarded as a a correction.

\subsection{String theory}

String theory already contains a minimum length as can be seen
from a simple argument \cite{ka90,kp90}: After one fixes the
reparametrization gauge, the partition function for a string
can be written in the discretized form
\[
Z=\int{\cal D} x\exp\left\{ i\frac{{\rm
const.}}{l_*^2}\sum\epsilon^2\left[\left(\frac{\delta_\tau
x}{\epsilon}\right)^2 +\left(\frac{\delta_\sigma
x}{\epsilon}\right)^2 \right]\right\},
\]
where the constant appearing in the exponent depends on the
particular model that one is considering. We can see that, due
to the fact that the worldsheet is two dimensional, the
$\epsilon$-factors cancel out, thus leading to an
$\epsilon$-independent spacetime distance
\[
\langle (\delta x)^2\rangle \sim {\rm const.}\ l_*^2
\]

Through a detailed analysis of thought-experiments involving
high energy string collisions \cite{ve86,ac89}, and also
through a renormalization group type analysis in string theory
\cite{kp90,nyo89,gk91} it was found that this result could be
qualitatively written in the form (see, for instance, Refs.
\cite{me92a,st91})
\[
\Delta x\gsim \frac{1}{\Delta p}+{\rm const.}\  l_*^2\Delta p,
\]
essentially the same expression that had been previously
obtained in microscope thought-experiments. The linear term is
due to the behaviour of strings at high energy: They spread
out, their size being proportional to their energy.

Duality is also involved in the lower bound to the spacetime
distance \cite{np87,bv89,nko87,nko91,nag93,nas94}. Since it is
not possible to distinguish a compactification of radius $r$
from another of radius $l_*^2/r$, very small distances are the
same as very large distances and thus Planck's length
represents an insurmountable border (see, for instance, Ref.
\cite{st91}).

On the other hand, Klebanov and Susskind \cite{ks88} have shown
that one can obtain an exact description of the bosonic
string theory by constructing a discrete field
theory. Therefore, the number of short-distance degrees of
freedom of the bosonic string must be much smaller than in any
conventional field theory. This may be one of the reasons for
the exponential damping that high-energy fixed-angle scattering
amplitudes suffer \cite{gm87,gm88}, and for the slower growth
of thermal partition functions at high temperature \cite{aw88}.
These are indications of the impossibility of resolving the
short-distance structure of string theory.

\subsection{Loop representation}

In the loop representation \cite{rs88,rs90,ro91a} (for a
review, see for instance Ref. \cite{sm92}) of non-perturbative
canonical quantum  gravity, geometry is studied
non-perturbatively, so that background fields do not play any
role. Operators that carry the metric information, but that are
made finite without the aid of any background structure, can be
defined. These operators can be used to define loop states
that, in the classical limit, provide a classical geometry. It
is remarkable that some of these finite operators have discrete
spectrum, i.e. a discrete small-distance structure. One such
operator is the area operator.

Let us consider a surface defined by a physical reference
system \cite{ro91}. A matter field can play such role, the
surface being the set of points where the field takes a
given value. Then, the area of this surface is quantized, i.e.
it is an integer multiple of the elementary Planck's area
\cite{ar92,ro93}. The integer number can be seen to be the
number of times that the loops cross the surface, so that each
loop contributes a unit (Planck's area) to the total area  when
it crosses the surface, regardless of the orientation of the
intersection.

One should note a difference between this approach and any
other in this work. Here, a small-distance scenario is built
and the consequences of this scenario are the recovery of a
classical geometry in the limit of large distances and the
realization that the small-scale structure is discrete. On the
other hand, the other analyses that we have studied here
approach the lower bound to the distance uncertainty from
above, that is from semiclassical reasonings concerning the
quantum fluctuations of the gravitational field.

\subsection{Black hole horizon}

It has been proposed by Bekenstein \cite{be74}  that the area
of a Kerr black hole should be quantized (see also Refs.
\cite{mu86,nko86,ga93b}). For simplicity, we will concentrate
on a Schwarzschild black hole. He argued that the area of a
black hole has a close resemblance to the action integral
$\oint pdx$ of a periodic mechanical system: first, both of
them are adiabatic invariants; and second, both of them have
dimensions of an action. Therefore, he proceeded by analogy and
stated that the area of a black hole should obey the equivalent
to the Bohr-Sommerfeld quantization rule:
\[
A=\frac{1}{2}l_*^2n,
\]
where n is an integer and the coefficient $1/2$ can be obtained
by means of the correspondence principle: for large $n$ the
system must have a classical behaviour. This result has been
reproduced from the semiclassical analysis of the
Wheeler-DeWitt equation for a collapsing dust star \cite{pe93}.
Bekenstein also realized that, except for the ground state
$n=1$, the Compton radius of the black hole $l_*^2/\sqrt A\sim
l_*/n$ is always smaller than its gravitational radius $\sqrt
A\sim l_*\sqrt n$. Only for a black hole in its quantum state
will the two  radii coincide, and be equal to Planck's
length.

Quite recently, Maggiore \cite{ma93a} has found that an
uncertainty relation that gives rise to a minimum length can
also be derived from the measurement of the radius of a black
hole. He realized that in classical general relativity the
observer does not have access to the horizon and the only way
of measuring its radius is through the relation between the
radius of the apparent horizon and the parameters of the black
hole (mass, charge, etc.). Therefore, this relation must be
regarded as a definition rather than as an experimentally
testable relation. The situation changes when quantum gravity
is put into the theory. Hawking radiation allows one to make an
independent, direct measurement of the area of the black hole
and, consequently, the relation between the radius of the
apparent horizon and the black hole parameters can be
experimentally tested (at least in thought experiments).

Let us perform one of these thought experiments by dropping a
photon of frequency $\omega$ into an extremal black hole. Then
the mass of the black hole will be incremented by an amount
$\omega$ and will decay to the extremal stable state. Assume
that this decay is accompanied by an emission of a single photon
of frequency $\omega$ (if more photons are emitted the same
conclussion will be reached). Note that this assumption is
plausible since the emission produced by a nearly extremal
black hole need not be thermal \cite{ps91}. By detecting this
photon and repeating the measurement many times, one can
``see'' the black hole. There is an uncertainty due to the
standard Heisenberg principle: $\Delta x\geq 1/\Delta p$ where
$\Delta p$ represents the uncertainty in the momentum of the
black hole, due to the emission of the photon, whose direction
is not well known. But also, since the mass of the black hole
will pass from $M+\omega$ to $M$ during the measurement, we can
only know the value of its mass with an uncertainty $\omega$.
Accordingly, the radius of the apparent horizon will have an
uncertainty $\Delta x\gsim l_*^2\omega\gsim l_*^2\Delta p$, so
that the uncertainty in the size of the hole is given by the
already familiar expression
\[
\Delta x\gsim {\rm max} \left(\frac{1}{\Delta p},l_*^2
\Delta p\right).
\]

\section{Ultraviolet convergence}

The lower bound on the proper distance between any two events
may become the ultimate ultraviolet regulator \cite{de57,le83},
since any distance below this bound loses its meaning. Let us
illustrate this issue  in the case of a massless scalar field
in curved spacetime \cite{de57,pa85a} (see also Ref.
\cite{pa88}). The Green function evaluated in a classical
spacetime has the well-known form
\[
D(x,y)\sim \frac{1}{l_0^2(x,y)+i\epsilon},
\]
which is divergent, i.e. goes to infinity for small proper
distances. However, when quantum fluctuations of spacetime are
taken into account, the notion of distance is no longer unique.
Therefore, an average over all possible distances provided by
the quantum fluctuations must be performed. This amounts to
replace the propagator $D(x,y)$ by its expectation value
\[
\langle D(x,y)\rangle\sim \frac{1}{l^2(x,y)+i\epsilon}\sim
\frac{1}{l_0^2(x,y)+l_*^2+i\epsilon}.
\]
We see that now the ultraviolet limit $x\rightarrow y$ gives a
finite propagator rather than a divergent one, because of the
existence of a minimum proper interval, which acts as a
high-energy cutoff. Indeed, by Fourier-transforming this
expression (and rotating it to Euclidean space) we obtain the
Euclidean propagator in momentum space
\[
\langle D(k)\rangle\sim \frac{l_*}{|k|}{\rm K}_1(l_*|k|).
\]
By asymptotically expanding the modified Bessel function ${\rm
K}_1(|k|)$, we can see how the presence of Planck's length
affects the propagator
\[
\langle D(k)\rangle\sim\left\{
\displaystyle
\begin{array}{ll}
1/k^2&\ \ \mbox{when}\ |k|\rightarrow 0\\
(l_*/|k|^3)^{1/2}e^{-l_*|k|}& \ \ \mbox{when}\
|k|\rightarrow \infty
\end{array}
\right. .
\]
It causes a strong damping on the high-energy contributions to
the propagator.

As expected, once the propagator has been made finite, one-loop
calculations show \cite{pa85a} that the effective potential is
also finite.

\section{Kinematical origin of the uncertainty relations}

In quantum mechanics, the uncertainty relations are not
dynamical. They come from the kinematical structure of the
theory, i.e. the fact that position and momentum operators
do not commute gives rise to an uncertainty relation between
them. The same will happen with any two observables that do not
commute: it will not be possible to measure them
simultaneously. Now we ask the question of whether the
generalized uncertainty relations can also be derived from the
kinematical structure of the theory. From the examples and
procedures that we have discussed, it seems quite clear that
the two pieces that contribute to the uncertainty in the
position are very different in nature. One of them  clearly has
a kinematical structure, stating  that objects are wavelike.
This part does not contain any dimensional parameter or
coupling constant. On the other hand, the second term involves
gravitation, which is a dynamical theory of
the interaction between matter and spacetime. That it contains
a dimensional constant, Planck's length, is a reflection of
this fact. Therefore, one can ask how it is possible to make
such  dynamically generated uncertainty relations a
consequence of some kinematical structure. The answer relies on
the fact that even though one could find it, it would be
necessary to give the dynamics of the system in order to
provide it with a clear well-defined meaning, since a suitable
non-local change of variables can make the uncertainty relation
look like Heisenberg's. However, this non-local change of
variables will turn the action non-local and non-local actions
contain a minimum length in them \cite{ka90}. In view of this
discussion, one cannot truly say that the generalized
uncertainty relations are due to kinematics exclusively, as is
the case with Heisenberg's principle. This discussion of the
kinematical or dynamical origin of the generalized uncertainty
relations may well be relevant to the final full quantization
of gravity since, as mentioned in the introduction, this seems
to be the origin of many problems when one is quantizing gravity:
the active response of spacetime to quantum fluctuations.

An early attempt to give a kinematical meaning to the
generalized uncertainty relations was made by Mead \cite{me66}
(see also Ref. \cite{go84}). His idea was to represent a
physical quantity $A$ such that $\Delta A\geq\alpha$ by an
operator that fails to commute with some Hermitian bounded
operator $Q$. Indeed, if $-1\leq Q\leq 1$ and $[A,Q]=i\alpha$,
for instance, then $\Delta A\ \Delta Q\geq\alpha$; since
$\Delta Q\leq 1$, we see that $A$ will satisfy the required
relation $\Delta A\geq \alpha$. Such an operator $A$ will be
called ``indeterminate'', following Mead. In ordinary quantum
mechanics, the momentum of a particle in a box is an
indeterminate operator, for example. Of course, in this
analysis, due care must be taken with boundary conditions and
domains of the operators. The operator $Q$ associated to the
indeterminate position operator in quantum gravity will contain
information about the  origin and the physical significance of
the fundamental length, i.e. about the quantum theory of gravity.

By considering gravity as the gauge theory of the deSitter
group, Townsend \cite{to77} introduced Planck's length into
gravity kinematically. The appearance of this dimensional
constant is due to the non-commutativity of the generators of
translations. Quantum gravity effects become relevant at the
same scale at which this non-commutativity is important.
Recently, Maggiore \cite{ma93b,ma93c} (see also Refs.
\cite{nke93,nkm94}) has pushed this kind of  arguments further
and has found an algebra that gives rise to the generalized
uncertainty relations and that, under quite general
assumptions, is essentially unique. This algebra turns out to
be a deformed Heisenberg algebra
\[
[x_i,x_j]=-il_*^2\epsilon_{ijk}J_k,
\ \ \ \
[x_i,p_j]=i\delta_{ij}\left( 1+l_*^2E^2\right)^{1/2},
\]
where $J_k$ are the generators of the three-dimensional
rotations and $E$ is the energy of the particle, $E^2=p^2+m^2$.
This result can also be obtained by considering the deformed
Poincar\'e algebra, where position operators are substituted by
suitably deformed Newton-Wigner operators \cite{ma93b}, rather
than the undeformed one.

There is an important difference between Maggiore's approach to
the existence of a minimum length and the other approaches that
we have considered so far. It is the issue of whether this
resolution limit also affects the time variable or not. In view
of the deformed Heisenberg algebra proposed by Maggiore, only
spatial separations will be subject to generalized uncertainty
relations. However, as we have seen in the approaches above,
the generalized uncertainty relations should apply to all
position and time measurements, the minimum uncertainty being
intrinsic also to synchronization of clocks. What there seems
to be subject to minimum uncertainty is the proper spacetime
interval  rather than the spatial separations alone.

\section{Measuring the gravitational field}

\subsection{Classical measurement devices}

The measurement process of the gravitational fields naturally
leads to a minimum volume of the measurement domain. Based on
the work by Bohr and Rosenfeld \cite{br33,ro63,br50,ro55} (see
e.g. Ref. \cite{he54})  for the electromagnetic field, Peres
and Rosen \cite{pr60} and then DeWitt \cite{de62} carefully
analyzed the measurement of the gravitational field and the
possible sources of uncertainty (see also Refs.
\cite{re58,bt82,tr85}). Their analysis was carried out in the
weak-field approximation (the magnitude of the Riemann tensor
remains small for any finite domain) although the features
under study can be seen to have more fundamental significance.
This approximation imposes a limitation on the bodies that
generate and suffer the gravitational field: their masses must
be small when compared with their linear dimensions. It is
worth noting that this minimum length does not appear in the
case of an electromagnetic field. The main reason for this is
that in this case, the relevant quantity that is involved in
the uncertainty relations is the ratio between the charge and
the mass of the test body, and this quantity can be made
arbitrarily small. This is certainly not the case for
gravitational interactions, since the equivalence principle
precisely fixes the corresponding ratio between gravitational
mass and inertial mass, and therefore it is not possible to
make it arbitrarily small. Let us go into more detail in the
comparison between the electromagnetic and the gravitational
fields as far as uncertainties in the measurement are concerned.

The arguments of Bohr and Rosenfeld lead to the uncertainty
relation
\[
\Delta F\ l^3\gsim \frac{q}{m},
\]
where $F$ is the electromagnetic field strength and $l$, $q$
and $m$ are the linear dimension, the charge and the mass of
the test body, respectively, provided that the test body
satisfies the conditions
\[
l\gsim \frac{1}{m},
\hspace{1cm}
l\gsim \frac{q^2}{m}.
\]
They are the reflection of the following assumptions concerning
the test body: the measurement of the field average over a
spacetime region, whose linear dimensions and time duration are
determined by $l$, is performed by determining the initial and
final momentum of a uniformly charged test body; the time
interval required for the momentum measurement is small
compared to $l$; any back-reaction can be neglected if the mass
of the test body is sufficiently high; and finally, the borders
of the test body are separated by a spacelike interval. These
conditions can be summarized by saying that the test body must be
classical from both the quantum and the relativistic point of
view. We see from the uncertainty relation for the
electromagnetic field that an infinite accuracy can be achieved
if an appropriate test body is used. This is not the case for
the gravitational interaction.  Indeed, the role of $F$ is now
played by $\Gamma/l_*$, where $\Gamma$ is the connection, and
the role of $q$ is played by $l_* m$. It is worth noting
\cite{pr60}, as mentioned above, that by virtue of the
equivalence principle, active gravitational mass, passive
gravitational mass and energy (rest mass in the Newtonian
limit) are all equal, and hence, for the gravitational
interaction, the ratio $q/m$ is the universal constant $l_*$.
The two requirements of Bohr and Rosenfeld are now
\[
l\gsim 1/m,
\hspace{1cm}
l\gsim l_*^2 m,
\]
so that $l\gsim l_*$. This means that the test body should not be
a black hole, i.e.  its size should not exceed its
gravitational radius, and that both its mass and linear
dimensions should  be larger than Planck's mass and length,
respectively. As in the electromagnetic case, Bohr and
Rosenfeld requirements can be simply stated as follows: the
test body must behave classically from the points of view of
quantum mechanics, special relativity and gravitation.
Otherwise, the interactions between the test body and the
object under study  would make this distinction (the test body on
the one hand and the system under study on the other) unclear as
happens in ordinary quantum mechanics: the measurement device
must be classical  or it it is useless as a measuring
apparatus. In this sense, in the context of quantum gravity,
Planck's scale stablishes the border between the measuring
device and the system that is being measured. Then, the
uncertainty relation for the connection can be written as
\[
\Delta \Gamma\ l^3\gsim l_*^2,
\]
or in terms of the metric tensor,
\[
\Delta g\ l^2\gsim l_*^2.
\]
The left hand side of this relation can be interpreted as the
uncertainty in the proper separation between the borders of the
region that we are measuring, so that it again states the
minimum position uncertainty relation. It is worth noting that
it is the concurrence of the three fundamental constants of
nature $\hbar$, $c$, and $\mbox{\small G}$ that leads to a
resolution  limit. If any of them is dropped then this
resolution limit disappears.

To end this section, a brief comment on the meaning of these
uncertainty relations is in order. DeWitt \cite{de62} and Peres
and Rosen \cite{pr60} interpreted them as providing a lower bound
to the size of the measurement domain and an indication of the
necessity of quantizing the gravitational field. On the other
hand, Rosenfeld \cite{ro63} and Borzeszkowski and Treder
\cite{bt88} regarded these uncertainties as an expression of the
impossibility of measuring quantum gravity effects, comparing
this situation with the controversy that the quantization of
the electromagnetic field raised in the early thirties
\cite{lp31,br33}.

\subsection{Continuous measurement}

Assume that we continuously measure an observable $Q$, within
the framework of ordinary quantum mechanics. Let us call
$\Delta q$  the uncertainty of our measurement device. This
means that, as a result of our measurement, we will obtain an
output $\alpha$ that will consist of the result $q(t)$ and any
other within the  range $(q-\Delta q, q+\Delta q)$. The
probability amplitude for an output $\alpha$ can be written
in  terms of path integrals \cite{me91}:
\[
A[\alpha]=\int_\alpha {\cal D} x e^{iS},
\]
where $\alpha$ denotes not only the output but also the set of
trajectories in configuration space that lead to it. For a
given uncertainty $\Delta q$, the set $\alpha$ is fully
characterized by its central value $q$. We are particularly
interested in studying the shape of the probability amplitude
$A$. More precisely, we will pay special attention to its width
$\Delta \alpha$ \cite{me91,me92b} (for a decoherent histories
approach see Ref. \cite{ha93}).

There are two different regimes of measurement, classical and
quantum, depending on whether the uncertainty of the measuring
device is large or small. If we define the set of classical
trajectories $\cal C$ as
\[
{\cal C}=\left\{ x\ /\ |S(x)-S(x_c)|\lsim 1\right\},
\]
then the classical regime of measurement will be accomplished
if $\Delta q$ is large enough so that ${\cal C}\subset
\alpha_c$, where $\alpha _c$ is the set of trajectories that
lie in the corridor defined by $q_c-\Delta q$ and $q_c+\Delta
q$. Since the set $\cal C$ is the one that gives significant
contribution to the path integral (the interference is
constructive), both the $\alpha_c$-restricted and the
unrestricted path integral will give about the same result. In
this regime, the width of the probability amplitude $\Delta
\alpha$ can be seen to be proportional to the uncertainty
$\Delta q$. Indeed, the greater $\Delta q$, the greater will be
the number of sets $\alpha$ that contain the set of classical
trajectories $\cal C$. Also, the uncertainty on the action can
be estimated to be
\[
\Delta S=|S(q_c+\Delta \alpha)-S(q_c)|\gsim|S(q_c+\Delta
q)-S(q_c)|\gsim 1,
\]
since $\Delta \alpha \gsim \Delta q$.

The quantum regime of measurement occurs when $\Delta q$ is so
small that no set $\alpha$ contains the set of classical
trajectories. Rather, it will happen that ${\cal C}\subset
\alpha_c$. Now the width of the probability amplitude is
$\Delta \alpha\sim 1/\Delta q$. Indeed, the sets $\alpha$ that
give a significant contribution to the path integral are those
that satisfy $|S(q+\Delta q)-S(q)|\lsim 1$. Therefore, the
smaller $\Delta q$ is, the easier will be this inequality to be
satisfied. The uncertainty in the action will be greater than
unity, because at least all the sets $\alpha$ contained in
$\cal C$ will have a high probability amplitude.

To summarize, in any regime of measurement, the action
uncertainty will be greater than unity. In view of the
discussion above, the width $\Delta \alpha$ of the probability
amplitude will achieve its minimum value, i.e. the measurement
will be optimized, for uncertainties in the measurement device
$\Delta q$ that are neither too large nor too small. When this
minimum non-vanishing value is achieved, the uncertainty in the
action is also minimized and set equal to one. The limitation
on the accuracy of any continuous measurement is, of course, an
expression of Heisenberg's uncertainty principle. Since we are
talking about measuring trajectories in some sense, a
resolution limit should appear, expressing the fact that
position and momentum cannot be measured simultaneously with
infinite accuracy. In the classical regime of measurement, the
accuracy is limited by the intrinsic uncertainty of the
measuring device. On the other hand, when very accurate devices
are employed, quantum fluctuations of the measuring apparatus
affect the measured system and the final accuracy is also
affected. The maximum accuracy  is obtained when there is achieved
a compromise between keeping the classical uncertainty low and
keeping quantum fluctuations also small.

This discussion is quite similar to the arguments that we have
made  throughout this work concerning the existence of a
minimum length, a resolution limit  in the context of quantum
gravity. In fact, any measurement of the gravitational field is
not only extended in time,
but also extended in space. These measurements are made by
determining the change in the momentum of a test body of a
given size. That measurements of the gravitational field have
to be extended in spacetime, i.e. they have to be
continuous, is due to the dynamical nature of this field.

Let us consider \cite{me92b} a measurement of the scalar
curvature averaged  over a spacetime region of linear dimension
$l$, given by the resolution of the measuring device (the test
body). The action is
\[
S=\frac{1}{l_*^2}\int d^4xR\sqrt{-g},
\]
where the integral is extended over the spacetime region under
consideration, so that it can be written  as
\[
S=\frac{1}{l_*^2}\bar R l^4,
\]
$\bar R$ being the average curvature. The action uncertainty
principle $\Delta S\gsim 1$  gives the uncertainty relation for
the curvature
\[
\Delta \bar R l^4\gsim l_*^2,
\]
equivalent to the previously obtained relations for the
connection and the metric.

We can see that the problem of measuring the gravitational
field, i.e. the structure of spacetime, can be traced back to
the fact that any such measurement is non-local, i.e. the
measurement device is aware of what is happening at different
points of spacetime and takes them into account. In other
words, the measurement device averages over a spacetime region.
The equivalence principle also plays a fundamental role: the
measurement device cannot decouple from the measured system and
back reaction is unavoidable.

\section{Small-distance structure}

\subsection{Saturation of Lorentz transformations}

We have seen that quantum, special relativistic and
gravitational effects, all together, give rise to a fundamental
length. Also, they may cause Lorentz transformations to saturate
at the Planck's scale. Indeed, let us apply a Lorentz
transformation characterized by the boost parameter $\gamma$ to
a particle of rest-mass $m$ and linear dimension $l$. For the
laboratory system of reference, its  linear dimension will be
$l/\gamma$, its mass $\gamma m$, its Compton wavelength
$1/\gamma m$ and its gravitational radius $l_*^2\gamma m$. Its
size will be given by the largest of these quantities. If the
boost is smaller than the critical  value
\[
\gamma_c=\frac{1}{l_*}{\rm max}\left(1/m,\sqrt{l/m}\right),
\]
then the size of the particle will be  determined by either
its linear dimension or its wavelength, whichever is larger,
and it will decrease with $1/\gamma$. Once the boost
parameter reaches its critical value,  the gravitational radius
will  define the size of the particle (assuming the absence
of other effects that would contribute to an increasing of its
size, such as those considered below)
and increase as $l_*^2\gamma
m$. Therefore, the particle will have a  minimum size no matter
how large the boost is. Although the critical value,
$\gamma_c$, of the parameter $\gamma$ for the saturated boost
depends on the mass and linear dimension  of the particle, the
minimum size that it may ever reach is Planck's length,
independent  of the nature of the particle. Of course, this
could be expected because if Lorentz transformations did not
saturate, then highly boosted particles could be used as
arbitrary accurate probes, in contradiction with the minimum
length uncertainty relation.

Due to the existence of a minimum proper interval, the notion
of event loses one of its characteristic features: it is no
longer an invariant concept \cite{su93b,su93a,ma93d}. The
uncertainty in the radius of the black hole horizon may be
interpreted as giving rise to a membrane, a stretched horizon,
from the point of view of an observer at rest with respect to
the black hole (for a study of the effect of a Planck's length
cutoff on black-hole evaporation, see also Ref. \cite{ja91}).
However, this membrane would be real only for this observer at
rest (its reality would be dramatically shown if this observer
were too close to the membrane since its extremely high temperature
would burn him to ashes). A free-falling observer would not see
or feel this membrane since, for him, nothing special would take
place at the horizon; in fact, there would be  no horizon for him.
The absolutely  different perceptions of both observers can be
stated \cite{st93b,st93a} (see also Refs. \cite{th85,th90} for
related ideas) in terms of a black hole complementarity
principle: events are observer-dependent, in the same manner as
in quantum mechanics the answer to the question ``Is an
electron particlelike or wavelike?''  depends on the kind of
measure that is performed, or as in special relativity the
issue of simultaneity of two events depends on the observer.
The physical laws appear  the same in all reference systems but
the description of the physical reality may vary from observer
to observer.

{}From the black hole complementarity principle, one can also
derive the saturation of Lorentz contraction when a particle
that is falling to a black hole acquires an energy close to
Planck's scale (as it reaches the stretched horizon). This
saturation results also in a minimum size for the particle but
of a slightly different nature from the one discussed above.
Here the notion of size that is involved is that of ``size
occupied by information'', which, as discussed by Susskind
\cite{su93a}, may not coincide with the wavelength or the
gravitational radius of the particle. According to the black
hole complementarity principle, the stretched horizon must
thermalize and spread throughout the information that falls to
the black hole, to be reemited afterwards. This means that, as
the particle reaches the stretched horizon (at that stage the
boost of the particle as seen by an observer at rest with
respect to the black hole is extremely large, close to
saturation), its longitudinal size must fully overlap with the
stretched horizon radial size, which is of the order of
Planck's length, and its transverse size must cover the whole
black hole in order to transfer all the information to the
stretched horizon homogeneously. Therefore, as the particle is
boosted, two different, simultaneous processes occur: first,
the longitudinal information size of the particle decreases up
to Planck's length and, second, its transverse information size
increases until it covers the stretched horizon.

\subsection{Causality}

The light cone itself is affected by the fluctuations of the
gravitational field. The uncertainty in the speed of photons
(the slope of the light cone) will increase with the
uncertainty in the gravitational field:
\[
\frac{\Delta v_{ph}}{v_{ph}}\sim\frac{\Delta g}{g}\gsim
\frac{l_*^2}{gl^2}.
\]
For fluctuations in the gravitational field of the order of the
gravitational field itself or, in other words, when the size of
the probe is close to Planck's scale, the uncertainty in the
light cone slope is as large as the slope itself. This means
that the distinction between spacelike and timelike separations
is lost close to  Planck's scale.

\subsection{Quantum cosmology}

In the quantum cosmology approach to quantum gravity (see e.g.
Ref. \cite{ha90}), wave functions are required to oscillate so
that they represent classical spacetime  when the radius of the
universe is larger than Planck's length, i.e.  they must have a
classical behaviour. However, these wave functions are also
well defined for small radii. For these values, at least in
simple models,  the wave function will represent a classically
forbidden region (it will no longer oscillate). This is nicely
illustrated by a simple minisuperspace model: de Sitter
spacetime. We will consider a homogeneous and isotropic
spacetime described by the Friedmann-Robertson-Walker metric
\[
ds^2=-dt^2+a(t)^2d\Omega_3^2,
\]
where $a(t)$ is the scale factor and $d\Omega_3^2$ is the
metric of the unit three-sphere. Then, the Wheeler-DeWitt
equation (up to operator ordering ambiguities) can be written as
\[
\left(-\partial_a^2+a^2-\lambda a^4\right) \Psi(a)=0,
\]
where $\lambda$ is the dimensionless cosmological constant and
$\Psi$ is the  wave function of the universe. For small $a$,
the potential term will be positive and therefore will
represent a classically forbidden region (tunnelling region).
In this simple example, the behaviour of the wave function is
\[
\Psi(a)\sim\left\{
\displaystyle
\begin{array}{ll}
\exp\left\{ \pm\displaystyle
\frac{1}{3\lambda}\left( 1-\lambda a^2\right)^{3/2}\right\} &
\ \mbox{when }\ a\ll 1/\sqrt\lambda\\
\exp\left\{ \pm\displaystyle
 \frac{i}{3\lambda}\left( \lambda a^2-1\right)^{3/2}\right\} &
\ \mbox{when }\ a\gg 1/\sqrt\lambda
\end{array}
\right. \ .
\]
It is oscillatory for large radii (classical behaviour) and
exponentially damped for small radii (classically forbidden
configurations). Therefore, the small radius region will
represent a genuine quantum spacetime, without a fixed metric,
causality relations or any other classical structure. In our
attempt of probing smaller and smaller distances, as the border
that separates both regions is reached, the quantum
fluctuations around the classical metric predicted by the
oscillatory regime become large. In fact, it is close to this
point where the semiclassical approximation breaks down.

\subsection{Topology change}

It has been suggested that the quantum fluctuations of the
metric would convert spacetime into a boiling magma in which
topology change is continuously happening
\cite{wh64,wh57,wh62,ha78}. This would mean that  time
evolution would no longer exist and causality would lose its
meaning, as has been seen to be the case due to the light cone
fluctuations. In view of these quantum fluctuations it is not
surprising that quantum tunnelling effects may take place. They
would cause particles to disappear from a region of spacetime and
reappear in another one. These tunnelling effects could be seen
as instantons, configurations in Euclidean space that would
connect, quantum mechanically, different regions of spacetime
otherwise disconnected or far apart from each other. They are
the so-called wormholes \cite{ha88} (for a review, see for
instance Refs. \cite{ha90b,ha90c,st89}). They admit also a full
quantum treatment in terms of wave functions
\cite{ha88,hp90,ga91,ga93}. The picture displayed above just
corresponds to the semiclassical approximation.  Actually, the
(Euclidean) path integral approach considers any possible
spacetime topology \cite{ha88,ha79}: the transition amplitude
between any two three-geometries can be written as the sum over
all possible topologies and over all possible metrics defined
on them that match the boundary data. Among these topology
changes, wormholes have deserved special attention. They
provide a scenario in  which the minimum length uncertainty
relation can be pictorially realized: wormholes would
delocalize any probe of sufficiently small size. This
nonlocality of the Planck's scale dynamics would be reflected
in the low-energy physics as effective interactions
\cite{ha88,co88,ks89,pr89} that would turn the coupling
constants of the probe, its mass/size among others, into
dynamical variables subject to a  probability distribution
\cite{co88}. This would be the low-energy remnant of the
Planck's scale uncertainties. Even more, it has been argued
that this probability distribution may have an infinite peak at
certain values, which would be the observed ones \cite{co88}.
This would imply that wormholes would dress the bare constants
of nature and therefore, when one is using a probe of certain
(low-energy dressed) mass/size, it would behave not as a
particle with these attributes but with the bare ones when
probing small distances, so that our measurement would have an
intrinsic uncertainty about small scale physics, this
uncertainty being greater and greater as the distance becomes
smaller and smaller.

\section{Conclusion}

Throughout this paper, we have displayed a series of arguments
in quantum gravity that  lead to the existence of a minimum
length, a resolution limit in any experiment. This means that
the notions of distance, of causality and any other notion
based on a metric structure lose their meaning at the Planck's
scale. This idea is very old and, in fact, can be traced back
to the first third of this century. What is remarkable is that
so many different analyses based on different ways of looking
at the problem of quantum gravity give the same prediction
about the nature of spacetime at the small-distance scale. The
presence of a lower bound to the uncertainty of distance
measurements seems to be a model-independent feature of quantum
gravity. It is a direct consequence of the physical laws that
affect the spacetime stucture. They can be summarized in the
three following well-know statements: $(i)$ the uncertainty
principle, $(ii)$ the speed of light is finite and constant,
and $(iii)$ the equivalence principle.

\acknowledgments

I am very grateful to Jonathan Halliwell and Max Ba\~nados for
fruitful discussions and suggestions and for a critical reading
of the manuscript. I also thank Alex Mikovi\'c and Roya Mohayaee
for useful conversations. The author was  supported by a joint
fellowship from the Ministerio de Educaci\'on y Ciencia (Spain)
and the British Council.

\end{document}